%% file: main.tex
\documentclass[runningheads]{llncs}
\usepackage[T1]{fontenc}
\usepackage{graphicx}
\usepackage{color}
\usepackage{hyperref}

\hypersetup{colorlinks=true,linkcolor=blue,citecolor=blue,urlcolor=blue}
\usepackage[ruled]{algorithm2e} % Algorithms
\usepackage{multirow}
\usepackage{enumitem}
\usepackage[misc]{ifsym}
\usepackage{booktabs}
\usepackage{amsmath}
\usepackage{cite}

\begin{document}
\title{A Universal Framework for Compressing Embeddings in CTR Prediction}
%
%\titlerunning{Abbreviated paper title}
% If the paper title is too long for the running head, you can set
% an abbreviated paper title here
%
\author{
Kefan Wang\inst{1} \and
Hao Wang\inst{1}\textsuperscript{(\Letter)} \and
Kenan Song\inst{2} \and
Wei Guo\inst{2} \and
Kai Cheng\inst{1} \and
Zhi Li\inst{3} \and
Yong Liu\inst{2} \and
Defu Lian\inst{1} \and
Enhong Chen\inst{1}\textsuperscript{(\Letter)}
}
\authorrunning{K. Wang et al.}
% First names are abbreviated in the running head.
% If there are more than two authors, 'et al.' is used.
%
\institute{
State Key Laboratory of Cognitive Intelligence, University of Science and Technology of China, Hefei, China\\
\email{\{wangkefan,ck2020\}@mail.ustc.edu.cn}\\
\email{\{wanghao3,liandefu,,cheneh\}@ustc.edu.cn}
\and
Huawei Singapore Research Center, Singapore\\
\email{\{songkenan,guowei67,liu.yong6\}@huawei.com}
\and
Shenzhen International Graduate School, Tsinghua University, Shenzhen, China\\
\email{zhilizl@sz.tsinghua.edu.cn}
}
% \institute{The ustc, Hefei, China \email{xxx@ustc.edu.cn}}
%
\maketitle              % typeset the header of the contribution
\input{Sections/0_abstract}
%
%
%
% \vspace{-2pt}
\input{Sections/1_intro}

\input{Sections/2_related_work}

% \vspace{-2pt}
\input{Sections/3_preliminary}

% \vspace{-2pt}
\input{Sections/4_method}

% \vspace{-2pt}
\input{Sections/5_experiment}

% \vspace{-2pt}
\input{Sections/6_conclusion}

% \subsubsection{\ackname} This work was supported by the National Natural Science Foundation of China (U23A20319, 62472394, 62441239, 62202443) as well as the Anhui Province Science and Technology Innovation Project (202423k09020011) and Anhui Provincial Science and Technology Major Project (No. 2023z020006)

\bibliographystyle{splncs04}
\bibliography{base}

% \appendix
% \section{Appendix}
% \input{Sections/7_appendix}
\end{document}

%% file: Sections/0_abstract.tex
\begin{abstract}

Accurate click-through rate (CTR) prediction is vital for online advertising and recommendation systems. Recent deep learning advancements have improved the ability to capture feature interactions and understand user interests. However, optimizing the embedding layer often remains overlooked. Embedding tables, which represent categorical and sequential features, can become excessively large, surpassing GPU memory limits and necessitating storage in CPU memory. This results in high memory consumption and increased latency due to frequent GPU-CPU data transfers. 
To tackle these challenges, we introduce a \underline{M}odel-agnostic \underline{E}mbedding \underline{C}ompression (MEC) framework that compresses embedding tables by quantizing pre-trained embeddings, without sacrificing recommendation quality. Our approach consists of two stages: first, we apply popularity-weighted regularization to balance code distribution between high- and low-frequency features. Then, we integrate a contrastive learning mechanism to ensure a uniform distribution of quantized codes, enhancing the distinctiveness of embeddings. Experiments on three datasets reveal that our method reduces memory usage by over 50x while maintaining or improving recommendation performance compared to existing models. The implementation code is accessible in our project repository {\color{blue} \url{https://github.com/USTC-StarTeam/MEC}}.

\end{abstract}

%% file: Sections/1_intro.tex
\section{Introduction}
Click-through rate (CTR) prediction is a pivotal task in online advertising and recommender systems, aimed at estimating the likelihood of a user clicking on a given item~\cite{other_18_clustering,other_16_user}. Traditional approaches such as Logistic Regression (LR) \cite{LR} % \hao{ref} 
and Factorization Machines (FM) \cite{FM}  % \hao{ref} 
model feature interactions to capture linear and pairwise relationships, effectively laying the groundwork for CTR prediction. However, these methods often fall short of capturing complex high-order interactions. With the fast development of deep learning~\cite{other_21_learning,other_22_context,other_23_graph,other_24_hierarchical,other_25_cglb,other_26_exploring,other_27_configure,other_28_lever,other_29_live}, some deep learning-based models have emerged to address these limitations, significantly improving the ability to capture intricate feature interactions~\cite{important_1_survey,important_2_perlaw,important_5_multi,other_6_breaking,other_7_bridging,other_11_exploring,other_13_efficient,other_14_mf}. Notable examples include DeepFM \cite{DeepFM}, which combines FM \cite{FM} with deep neural networks, and Product-based Neural Networks (PNN) \cite{PNN}, which explicitly model feature interactions through product layers. Further advancements like Adaptive Factorization Networks (AFN) \cite{AFN} and Deep \& Cross Networks (DCN) \cite{DeepCross} introduce adaptive mechanisms and innovative cross-layer structures. Recently, Gated Deep \& Cross Networks (GDCN) \cite{GDCN} have been proposed to offer greater flexibility and improved performance. Additionally, user modeling-based CTR models such as Deep Interest Network (DIN) \cite{DIN} and Deep Interest Evolution Network (DIEN) \cite{DIEN} further enhance CTR prediction by capturing user interests and their evolution over time.

    Despite advancements in CTR prediction, embedding tables can become exceedingly large~\cite{others_1_scaling,other_10_entropy,others_2_td3}, often reaching hundreds of gigabytes in industrial settings \cite{QRTrick,BinaryCode,important_3_scaling}. 
    This substantial memory consumption frequently exceeds GPU capacity, causing data transfers between GPU and CPU during inference, which impacts performance and introduces latency in real-time systems \cite{wang2022merlin}. 
    To address this, memory-efficient techniques have been proposed, generally falling into hash-based and quantization-based methods. Hash-based methods, such as DHE \cite{DHE}, DoubleHash \cite{DoubleHash}, and QRTrick \cite{QRTrick}, reduce memory usage by mapping features to fewer buckets using hash functions. 
    However, they suffer from collision issues that degrade performance by confusing unrelated concepts.

    \begin{figure}[t]
        \centering
        \setlength{\abovecaptionskip}{5pt}   % 图注上方的距离
        \setlength{\belowcaptionskip}{0pt}   % 图注下方的距离
        \includegraphics[width=0.9\textwidth]{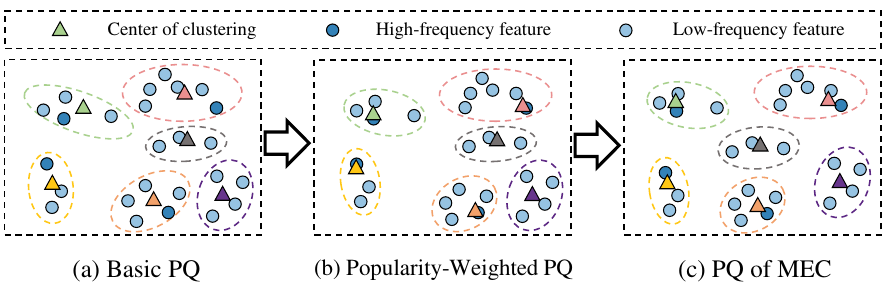}
        \caption{Distribution of different quantization methods}
    \label{fig:motivation}
    % \vspace{-15pt}
    \end{figure}

    In contrast, quantization methods like Product Quantization (PQ) \cite{PQ} offer better performance with reduced memory consumption by decomposing high-dimensional embeddings into subspaces and quantizing them separately~\cite{other_17_guesr}. 
    Recent advancements include DPQ~\cite{DPQ}, AutoDPQ~\cite{AutoDPQ}, and CCE\cite{CCE}. 
    Yet, these methods often overlook the quality of quantized representation distribution, leading to suboptimal results due to \textsl{\textbf{imbalances in code allocation}} and \textsl{\textbf{uneven distributions of code embeddings}}. 
    Frequency-independent quantization can be disrupted by low-frequency features, overshadowing high-frequency ones and weakening representation, as shown in Fig. \ref{fig:motivation}(a). 
    Moreover, directly considering frequency might overwhelm low-frequency features, as depicted in Fig. \ref{fig:motivation}(b). 
    Additionally, CTR tasks require evenly distributed embeddings to maintain distinctiveness, but traditional methods often lead to concentrated embeddings, compromising predictive performance \cite{letter}.

    To address these challenges, we propose a Model-agnostic Embedding Compression (MEC) Framework. 
    Firstly, to tackle \textbf{imbalances in code allocation}, we introduce popularity-weighted regularization (PWR). PWR preserves high-frequency feature information, preventing it from being overshadowed by low-frequency features. 
    It penalizes imbalances, ensuring low-frequency features aren't dominated by high-frequency ones. 
    By adaptively assigning unique cluster centers to high-frequency features while promoting code-sharing among low-frequency features, PWR enhances representation capacity. 
    Secondly, to mitigate \textbf{uneven distributions of code embeddings}, we integrate a contrastive learning mechanism.  
    Using random code replacement to generate semi-synthetic negatives, the contrastive loss encourages uniform distribution of codes, ensuring distinctiveness. 
    As shown in Fig. \ref{fig:motivation} (c), this method enhances recommendation performance by providing more accurate and diverse representations.

    Our proposed framework operates in two stages, which enhances its generalization capabilities and allows for easy integration with any state-of-the-art quantization model or CTR model. 
    We instantiated the MEC algorithm on three representative CTR models and conducted extensive experiments on three datasets. The results demonstrate that MEC can adaptively perform high-quality quantization for data with varying popularity levels, achieving comparable or even superior results to existing advanced CTR models while saving more than 50x the memory.
    In summary, the contributions of this work are as follows:
    \begin{itemize}[leftmargin=*,align=left]
        \item We successfully address imbalances in code allocation by leveraging data popularity distribution through an adaptive two-stage quantization-based CTR prediction method, thereby enhancing recommendation performance while significantly reducing memory usage.
        \item To tackle the challenging issue of uneven distributions of code embeddings in PQ quantization for CTR tasks, we introduce PWR and contrastive learning methods to significantly enhance the quality of quantized codes, thereby improving recommendation performance.
        \item Extensive experiments on three real-world datasets validate that our MEC framework achieves over 50x memory optimization compared to models with conventional embedding layers, surpassing baselines in both recommendation performance and memory usage.
    \end{itemize}

%% file: Sections/2_related_work.tex
\section{Related Work}
    \subsection{Click-Through-Rate Prediction}
        CTR prediction focuses on estimating the likelihood of user clicks. Existing methods are mainly divided into feature interaction-based and user behavior modeling-based approaches. Feature interaction models, such as Wide\&Deep \cite{WideDeep} and DeepFM \cite{DeepFM}, utilize network architectures to capture complex interactions. For example, GDCN \cite{GDCN} employs dual tower networks to model both explicit and implicit interactions.

        User behavior modeling aims to extract personal preferences from historical interactions~\cite{important_2_perlaw,important_4_dataset,other_3_fuxi,other_12_learning,other_15_apgl4sr,other_19_hypersorec,other_20_mcne}. Techniques have evolved from attention networks \cite{zhou2018deep} to efficient Transformers. DIN \cite{zhou2018deep} uses attention to assign scores to user behaviors, capturing diverse interests.

    \subsection{Memory-Efficient Recommendation}
        Memory-efficient techniques, including hashing and quantization, are crucial for resource-constrained systems. Hashing methods convert embeddings into compact codes, improving memory efficiency but risking noise from unrelated mappings \cite{hashing_trick,DHE}. Quantization reduces memory usage by representing embeddings through codebooks, achieving higher compression rates \cite{DPQ}. Methods like AutoDPQ adaptively determine codebook sizes, while CCE combines PQ and hashing for dynamic updates \cite{AutoDPQ}.

        However, these approaches often overlook the distribution quality of quantized embeddings. Our adaptive two-stage training framework optimizes embedding distributions post-quantization, enhancing representation quality without significant computational overhead. This approach complements existing methods and offers potential for future research advancements.

%% file: Sections/3_preliminary.tex
% \section{Click-Through Rate (CTR) Prediction}
\section{Preliminary}

The CTR prediction task is formulated as predicting the probability $\hat{y} = P(\text{click} | \mathbf{x})$ of a user clicking based on input features $\mathbf{x}$, which is crucial for online recommendation platforms. 
The input features can be divided into two types: categorical features and numerical features.
For categorical features, we start from an embedding layer that transforms the raw features of a training sample into embedding vectors. 
For a feature $A$ with vocabulary size $v_A$, we first encode the categorical information into a one-hot or multi-hot vector $\mathbf{x}_{\text{A}} \in \{0,1\}^{v_A}$, then an embedding lookup operation is conducted to transform the high-dimensional feature vector into a low-dimensional embedding:
\begin{equation}
\mathbf{e}_\text{A} = \mathbf{E}_\text{A} \mathbf{x}_\text{A},
\end{equation}
where $\mathbf{E}_\text{A} \in \mathbf{R}^{d_A \times v_A}$ is the embedding table of feature $A$.
Noticed that for sequential multi-hot categorical features here, we apply the average pooling here to transform it into a single vector.
For numerical features, each one is processed through a DNN layer to map the raw value into a vector. Let $\mathbf{x}_{a}$ be the $a$-th continuous feature, transformed as follows:
\begin{equation}
\mathbf{e}_{a} = \text{DNN}_a(\mathbf{x}_{a}),
\end{equation}
where $\text{DNN}_a$ is the layer specific to the $a$-th feature.
By concatenating all features, we represent an instance as:
\begin{equation}
\mathbf{e} = [\mathbf{e}_{1}, \mathbf{e}_{2}, \ldots, \mathbf{e}_{N}],
\end{equation}
with $N$ being the total number of features. The combined embedding vectors are then input into the CTR prediction model to capture complex interactions and predict click probabilities.

%% file: Sections/4_method.tex
    \begin{figure*}[ht]
        \centering
        \setlength{\abovecaptionskip}{5pt}   % 图注上方的距离
        \setlength{\belowcaptionskip}{-6pt}   % 图注下方的距离
        \includegraphics[width=\textwidth]{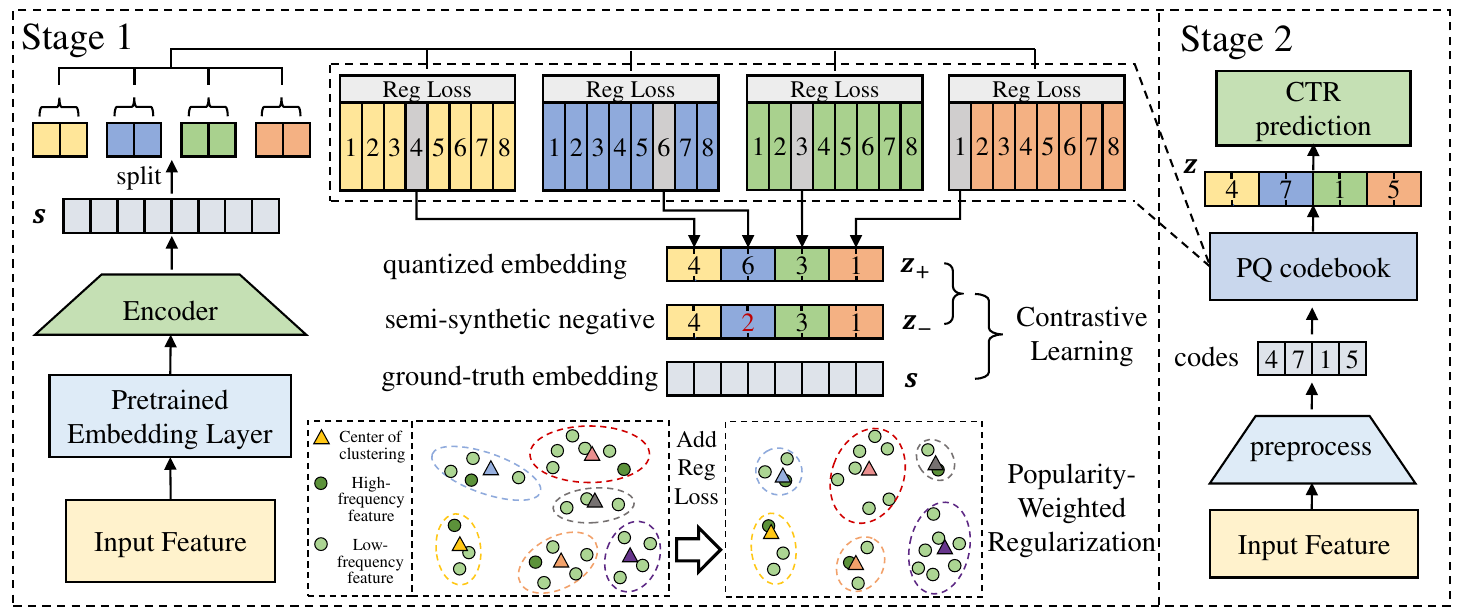}
        \caption{Overview of our MEC framework. The framework consists of two stages: pre-training and downstream task training. In the first stage (left), a PQ codebook is learned by combining existing embeddings. In the second stage (right), the input features are quantized based on the PQ codebook and used to train a CTR model for the downstream task. During online inference, the pre-quantized features are fed into the downstream task to achieve memory-efficient CTR prediction.}
    \label{fig:main}
    \vspace{1em}
    \end{figure*}
\section{Methodology}

Contemporary CTR prediction methods face challenges with large embedding spaces, leading to increased storage demands and reduced inference efficiency. Traditional quantization techniques often overlook high-frequency features and fail to evenly distribute embeddings, degrading model performance. Our approach compresses embedding spaces while maintaining accuracy, as detailed in Section 4.1 for the MEC framework, Section 4.2 for popularity-weighted regularization (PWR), and Section 4.3 for contrastive learning to address uneven representation distribution.

\subsection{Overview}
Existing models typically rely on dense embeddings, resulting in large tables and slower inference. While quantization is a common solution, it must effectively retain feature richness for accurate CTR prediction. Current models are not fully optimized for this, adversely affecting recommendation performance. As shown in Fig. \ref{fig:main}, our innovative two-stage framework decouples quantization from inference, allowing easy adaptation and efficient updates.

In the first stage, an auxiliary CTR model pre-trains embeddings. The input features $\mathbf{x}$ are transformed into embedding vectors $\mathbf{e}$ via an embedding function $F_{\text{Emb}}$, then encoded and partitioned for quantization:

\begin{equation}  
\mathbf{e} = F_{\text{Emb}}(\mathbf{x}), \quad \mathbf{s} = \text{Encoder}(\mathbf{e}).
\end{equation}  

Here, $\mathbf{x}$ represents the input features, $\mathbf{e}$ is the resulting embedding vector, and $\mathbf{s}$ is the encoded vector prepared for quantization. The vector $\mathbf{s}$ is split into $M$ sub-vectors $\{\mathbf{s}_1, \mathbf{s}_2, \ldots, \mathbf{s}_M\}$, each of dimension $d/M$, where $d$ is the dimension of the encoded vector. For Product Quantization (PQ), each sub-vector $\mathbf{s}_i$ is quantized using a codebook $\mathbf{C}_i$, initialized with size $K$. The closest codeword is identified by minimizing the Euclidean distance:

\begin{equation}  
j_i = \arg \min_j \| \mathbf{s}_i - \mathbf{c}_{i,j} \|_2, \quad \mathbf{q}_i = \mathbf{c}_{i,j_i},
\end{equation}  

where $\mathbf{c}_{i,j}$ is the $j$-th codeword in the codebook $\mathbf{C}_i$. The quantized vector $\mathbf{q}$ is formed by concatenating all $\mathbf{q}_i$:

\begin{equation}  
\mathbf{q} = [\mathbf{q}_{1}, \mathbf{q}_{2}, \ldots, \mathbf{q}_{M}].
\end{equation}  

To ensure $\mathbf{q}$ closely approximates $\mathbf{s}$, we minimize the reconstruction loss:

\begin{equation}  
\mathcal{L}_{\text{recon}} = \| \mathbf{s} - \mathbf{q} \|_2^2.
\end{equation}  

After convergence, the learned codebook $\mathbf{C}$ and quantization method $\phi$ are retained for CTR prediction. Enhancements include Popularity-weighted Regularization and Contrastive Product Quantization, which improve the distribution of quantized features.

\subsubsection{Stage 2: Memory-Efficient Embedding for CTR Prediction}
Using the saved quantization method $\phi$, input features $\mathbf{x}$ are transformed into quantized codes $\mathbf{c}$, retrieving embeddings for CTR model training and inference:

\begin{equation}
\mathbf{c} = \phi(\mathbf{x}), \quad \mathbf{e} = \mathbf{E}_{\phi} \mathbf{c},
\end{equation}

where $\mathbf{E}_{\phi}$ is initialized from the retained codebook $\mathbf{C}$. The CTR model predicts click-through rate probability using the embedding vector $\mathbf{e}$:

\begin{equation}
\hat{y} = \psi(\mathbf{e}),
\end{equation}

where $\psi$ represents the feature interaction operations in the downstream CTR model. The prediction is optimized using a binary classification loss:

\begin{equation}
\mathcal{L}_{\text{binary}} = - (y\log(\hat{y}) + (1-y)\log(1-\hat{y})),
\end{equation}

where $y$ is the true label. By compressing the embedding table with the PQ codebook, we address memory consumption and enhance model inference efficiency and generalization.

    \subsection{Popularity-Weighted Regularization}
        \label{subsec:reg}

        In quantized CTR recommendation tasks, models often focus on memory efficiency, which may lead to imbalanced code allocation and overshadow critical high-frequency features. 
        Methods ignoring frequency fail to account for varying feature importance, causing representation issues. 
        However, introducing feature frequency directly usually overwhelms low-frequency features, leading to degraded performance. 
        
        To address these issues, we propose a popularity-weighted regularization method. We weigh features based on popularity, using a logarithmic transformation to smooth frequency disparities and calculate weights more effectively.
        The formula is as follows:
        \begin{equation}
        r_j = \lfloor \log_2 (n_j) \rfloor
        \end{equation}
        where $n_j$ denotes the frequency count of feature $j$.
        This allows the quantization model to adaptively distinguish features of different popularities. 
        However, this also causes the cluster centers to shift too much towards high-frequency features, making it difficult to effectively model low-frequency features. This is because, during the weighting process, although the weights of high-frequency features are smoothed, they still occupy a large proportion, thereby affecting the clustering effect of low-frequency features.

        To enhance the learning of low-frequency features, it is essential to incorporate regularization to balance the influences of high-frequency and low-frequency features. 
        We propose a loss function that integrates a regularization term to penalize imbalances in code allocation. 
        Given the efficacy of entropy-based tests in detecting uniformity \cite{dudewicz1981entropy}, we employ a modified entropy-based metric as the loss function to evaluate the uniformity of code distribution. This metric is selected for its capacity to measure the deviation from a uniform distribution, thereby facilitating fair code assignments.
        The regularization loss is calculated as follows:
        \begin{equation}
        \mathcal{L}_{\text{reg}} = \exp\left(-\sum_{i=1}^{K} p_i \log(p_i + \epsilon)\right), p_i = \frac{\sum_{j \in S_i} r_j}{\sum_{j \in S} r_j}
        \label{eq:loss_reg}
        \end{equation}
        where $K$ is the number of codes (embeddings), $\epsilon$ is a very small number (e.g., $1e-10$) to prevent the logarithm from being zero, $p_i$ is the probability of the $i$-th code, $S_i$ is the set of features assigned to code $i$, $S$ is the set of all features, and $r_j$ represents the popularity weighting of feature $j$. This calculation ensures that $p_i$ reflects the proportion of total feature frequency assigned to code $i$.
        Through this popularity-weighted regularization method, we can better balance the modeling of high-frequency and low-frequency features, thereby improving the overall performance.
        
        In summary, the popularity-weighted regularization method weights and smooths samples, allowing the quantization model to better handle features with different frequencies, avoiding excessive bias towards high-frequency features and neglect of low-frequency features. This method not only improves the accuracy of the model but also enhances its robustness and generalization ability.
        
    \subsection{Contrastive Product Quantization}
        \label{subsec:cons}
        
        While PQ embeddings and popularity-weighted regularization address code allocation imbalances, prior research \cite{letter} highlights that quantized embeddings often suffer from uneven distributions, leading to homogenization and reduced diversity.
        This lack of diversity hampers the model's ability to differentiate features, compromising CTR prediction accuracy. To address this, we integrate contrastive learning to enhance embedding diversity, ensuring balanced code allocations and improving recommendation performance.
        
        In contrastive learning, informative negative samples are crucial. We synthesize enhanced feature indices as negatives. However, fully synthetic indices may be too distant from real features. Thus, we create semi-synthetic codes based on real item codes to serve as effective hard negatives.
        
        Given a real feature code $\mathbf{c} = [j_1, j_2, \ldots, j_M]$, we randomly replace one index with a probability $\rho \in (0, 1)$ while keeping all others unchanged. This effective technique ensures the semi-synthetic codes resemble real features but still offer enough variability to act as challenging hard negative samples. The semi-synthetic code $\mathbf{z}$ is generated as follows:
        \begin{equation}
        G(\mathbf{z_i}) = 
        \begin{cases} 
        \text{Uni}(\{1, \ldots, K\}), & \text{if } X = 1 \\
        \mathbf{c_i}, & \text{if } X = 0
        \end{cases}
        \end{equation}
        where $X \sim \text{Bernoulli}(\rho)$, and $\text{Uni}(\cdot)$ uniformly samples item codes from the input set. This uniform sampling method guarantees that the code distribution of semi-synthetic indices is similar to that of real indices. The embedding of the real feature code $\mathbf{c}'$ and the corresponding semi-synthetic hard negative sample instance $\mathbf{z}'$ is given by:
        \begin{equation}
        \mathbf{c}' = \text{Emb-Pool}(\mathbf{c}), \mathbf{z}' = \text{Emb-Pool}(\mathbf{z}),
        \end{equation}
        where $\text{Emb-Pool}(\cdot)$ denotes the embedding lookup and aggregation. We use concatenation for aggregation here for simplicity.
    
        To incorporate contrastive learning into our framework, we define a contrastive loss function that encourages the model to distinguish between real item codes and their semi-synthetic hard negative samples. The contrastive loss $\mathcal{L}_{\text{con}}$ is defined as follows:
    
        \begin{equation}
        \mathcal{L}_{\text{con}} = -\frac{1}{N} \sum_{i=1}^{N} \log \frac{\exp(\text{sim}(\mathbf{s}, \mathbf{c}'))}{\exp(\text{sim}(\mathbf{s}, \mathbf{c}')) + \sum_{j=1}^{K} \exp(\text{sim}(\mathbf{s}, \mathbf{z}'))},
        \label{eq:loss_con}
        \end{equation}
        where $N$ is the number of features, $\text{sim}(\cdot, \cdot)$ denotes a similarity function (e.g., cosine similarity), 
        $\mathbf{s}$ is the encoded vector,
        $\mathbf{c}'$ is the quantized embedding vector, and $\mathbf{z}'$ are the quantized embeddings of $K$ semi-synthetic hard negative samples.
        By minimizing this contrastive loss, the model learns to bring the embeddings of the encoded vector and quantized vector closer together while pushing the semi-synthetic hard negative samples further apart. This results in a more balanced and diverse code embedding distribution.

%% file: Sections/5_experiment.tex
\section{EXPERIMENTS}

    %  We conduct extensive experiments on three real-world datasets (\gw{including two public accessible datasets and one private industrial dataset}) to answer the following research questions:
    % \begin{itemize}
    %     % \item \textbf{(RQ1)} 我们的模型与sota的CTR模型与量化推荐模型相比效果如何？（对应我们的overall performance部分）
    %     \item \textbf{(RQ1)} How does MemE-CTR perform compared to state-of-the-art CTR models and quantized recommendation models?
    
    %     % \item \textbf{(RQ2)} 我们的模型的不同模块如何带来提升？（对应我们的ablation study部分）
    %     \item \textbf{(RQ2)} How do the different modules of MemE-CTR contribute to performance improvement?
    
    %     % \item \textbf{(RQ3)} 我们的模型对实验设置的变化是否敏感，是否会明显受到超参数或预训练模型改变的影响？（对应我的Hyper-Parameter Analysis和Pretraining method analysis）
    %     \item \textbf{(RQ3)} Is our model responsive to variations in the experimental setup, and is it significantly influenced by alterations in hyperparameters or the pre-trained model?
    %     % \gw{Maybe put the industrial datasets experiments on a single section?} \kf{OK.}
        
    %     % \item \textbf{(RQ4)} 我们的模型在节约内存和提升速度方面能够带来什么样的提升？（对应我们的Time\&Memory Efficiency部分）
    %     \item \textbf{(RQ4)} What improvements can MemE-CTR bring in terms of memory efficiency and speed enhancement?

    %     % \item \textbf{(RQ5)} 为什么MemE-CTR要使用PQ作为它的量化方法，而不使用RQ等方法？
    %     \item \textbf{(RQ5)} Why does MemE-CTR use PQ as its quantization method instead of other methods?

    %     % RQ6 我们的模型能否在大规模工业数据集上同样取得优秀的表现？
    %     \item \textbf{(RQ6)} Can our model achieve high performance on large-scale industrial datasets?
    % \end{itemize}

    \input{Sections/Tables/dataset}

\subsection{Experiment Setup}
    \subsubsection{Datasets.}
        We evaluated our model using three datasets: two public and one private industrial dataset. 
        \begin{itemize}[leftmargin=*,align=left]
            \item Criteo: A standard dataset from Kaggle\footnote{https://www.kaggle.com/c/criteo-display-ad-challenge/} with one week of user click data for CTR prediction. It includes 45 million samples and 39 features (13 continuous, 26 categorical), useful for model evaluation.

            \item Avazu: Another Kaggle dataset\footnote{https://www.kaggle.com/c/avazu-ctr-prediction/}, used for CTR prediction with 11 days of user click data. It contains about 40 million samples and 24 features, serving as a solid base for testing algorithms.

            \item Industrial: A private dataset from the Huawei ad platform with over 400 million user impressions. It includes hundreds of features, both categorical and numerical, with 32 features having vocab sizes over 100,000, and the largest reaching six million.
        \end{itemize}
        We followed AFN \cite{AFN} preprocessing, splitting Criteo and Avazu datasets into training, validation, and test sets in a 7:2:1 ratio by time order. The Industrial dataset was split into train/val/test sets in a 6:1:1 ratio. Table \ref{tab:dataset} provides detailed statistics for the public datasets.

    \subsubsection{Baseline Models}
    To demonstrate the effectiveness of the proposed model, we select some representative CTR models for comparison.
    We also compare our proposed model with some hashing and quantization-based methods to validate the superiority of our model. 
    The details are listed as follows:
         
\textbf{Representative CTR Models:}
    \begin{itemize}[leftmargin=*,align=left]
        \item \textbf{LR} \cite{LR}: Logistic Regression is a linear model using a logistic function for binary variables. It is often a baseline in recommendation systems due to its simplicity and interpretability.
    % \end{itemize}     
    % \textbf{Factorization-based Models}
    % \begin{itemize}[leftmargin=*,align=left]
        \item \textbf{FM} \cite{FM}: Factorization Machine models pairwise interactions between features efficiently, making it suitable for sparse data.
        \item \textbf{AFM} \cite{AFM}: Attentional Factorization Machine enhances FM by using attention to model feature interaction importance.
        \item \textbf{DeepFM} \cite{DeepFM}: Combines FM for low-order and DNN for high-order interactions, improving recommendation accuracy.
    % \end{itemize}          
    % \textbf{Neural Network Models}
    %     \begin{itemize}[leftmargin=*,align=left]
            \item \textbf{PNN} \cite{PNN}: Product-based Neural Network uses product operations between feature embeddings to model feature interactions.
            \item \textbf{DCNv2} \cite{DCNv2}: Deep \& Cross Network v2 captures explicit and implicit feature interactions by stacking cross and deep layers.
            \item \textbf{AutoInt}\cite{AutoInt}: Utilizes self-attention mechanisms to automatically learn feature interactions without manual feature engineering.
            \item \textbf{AFN} \cite{AFN}: Adaptive Factorization Network dynamically learns the importance of feature interactions using a neural network.
            \item \textbf{SAM} \cite{SAM}: Self-Attentive Model leverages self-attention to capture complex feature interactions effectively.
            \item \textbf{GDCN} \cite{GDCN}: Gate-based Deep Cross Network uses gating mechanisms to model the complex relationships between users and items, enhancing collaborative filtering signals.
        \end{itemize}
              
    \textbf{Hashing and Quantization Models:}
        \begin{itemize}[leftmargin=*,align=left]
            \item \textbf{DHE} \cite{DHE}: Deep Hash Embedding uses hashing techniques for dimensionality reduction, improving memory and time efficiency.
            \item \textbf{xLightFM} \cite{xLightFM}: An extremely memory-efficient Factorization Machine that uses codebooks for embedding composition and adapts codebook size with neural architecture search.
        \end{itemize}
                % \gw{Where is AFM, AFN, SAM, DHE, and LightFM?}
    % \subsubsection{Evaluation Metrics.}
        \input{Sections/Tables/main_result} 

\subsubsection{Evaluation Metrics}
% We assess the algorithms using AUC and Logloss metrics. AUC (Area Under the ROC Curve) evaluates the model's ability to rank positive instances higher than negatives, with higher values indicating better performance. Logloss measures the discrepancy between predicted probabilities and actual labels, where lower values indicate better accuracy. Our method achieves over 90\% memory savings, maintaining competitive performance with baselines and offering significant practical value.
We evaluate the algorithms using AUC and Logloss metrics to ensure a comprehensive performance assessment. AUC (Area Under the ROC Curve) evaluates the model's capability to rank positive instances above negatives, with higher values indicating superior discrimination ability. Logloss, on the other hand, measures the accuracy of predicted probabilities in relation to actual labels, with lower values indicating better model calibration and precision. Our method achieves over 90\% memory savings while maintaining performance on par with baseline models. This efficiency, coupled with competitive accuracy, underscores the method's practical value, especially in environments where memory constraints are a concern.

\subsubsection{Parameter Settings}
All models were implemented using the FuxiCTR library\footnote{https://fuxictr.github.io/}. We standardized the embedding dimension to 40 and batch size to 10,000. The learning rate was chosen from \{1e-1, 1e-2, 1e-3, 1e-4\}, with $L_2$ regularization from \{0, 1e-1, 1e-2, 1e-3, 1e-4, 1e-5\}, and dropout ratios from 0 to 0.9. Models were trained using the Adam optimizer. Codebook sizes were \{256, 512, 1024, 2048\} and PQ layers \{2, 4, 8\}. Experiments were repeated five times to ensure reliability, reporting average results. Memory usage was based on full model size, and results are presented with optimal parameters. We used the pre-trained embedding layer of DeepFM as initialization before quantization.

    \subsection{Performance Comparison}
        % \subsubsection{Public Dataset Performance}
        % \label{subsec:public_performance}
        % In this section, we evaluate the performance of our proposed framework, MemE-CTR, on two widely-used CTR prediction datasets: Criteo and Avazu. 
        % In this section, we compare MemE-CTR with baseline models, and the results are presented in Table \ref{tab:MainTable}.
        % In this section, we compare MEC with baseline models in terms of both CTR performance and memory usage, and the results are presented in Table \ref{tab:MainTable}. We also conduct Wilcoxon signed rank tests \cite{p_test} to evaluate the statistical significance of MEC with the base model. We have the following observations:
        % The results in Table \ref{tab:MainTable} highlight several key findings:

        % \textbf{Finding 1:} Meme-CTR achieves performance that is on par with or superior to traditional CTR models. \gw{which model? on par with or superior? give a definite conclusion} 

        % \textbf{(1) The embedding tables play a great part in model parameters.} 
        % The parameters of traditional CTR models are mainly concentrated in the embedding layer. 
        % Despite the model structure and complexity varying a lot from each other, the overall parameters are much of the same size. 
        % As shown in Table \ref{tab:MainTable}, optimizing the embedding tables can reduce model parameters by over 90\%, highlighting the significant potential for quantization and compression techniques to improve efficiency and performance.
        
        \begin{figure}[t]
            \centering
            \setlength{\abovecaptionskip}{-5pt}   % 图注上方的距离
            \setlength{\belowcaptionskip}{0pt}   % 图注下方的距离
            \includegraphics[width=0.75\textwidth]{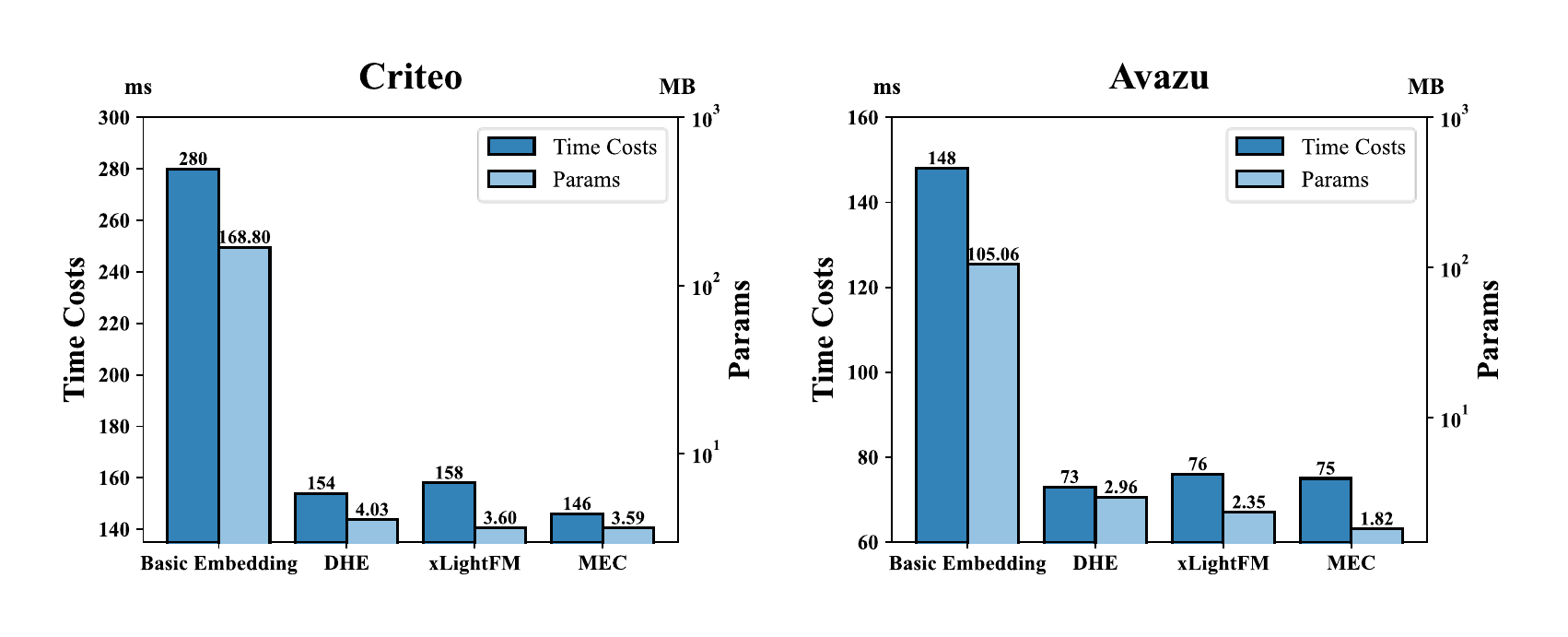}
            \caption{Time efficiency performance}
        \label{fig:time_eff}
        % \vspace{-14pt}
        \end{figure}

        In this section, we compare MEC with baseline models in terms of both CTR performance and memory usage, and the results are presented in Table \ref{tab:MainTable}. We also conduct Wilcoxon signed rank tests \cite{p_test} to evaluate the statistical significance of MEC with the base model. We have the following observations:
        
\textbf{(1) Embedding tables significantly impact model parameters.} 
In traditional CTR models, most parameters are concentrated in the embedding layer. Despite variations in model structure and complexity, the overall parameter size remains similar. Table \ref{tab:MainTable} shows that optimizing embedding tables can reduce model parameters by over 90\%, highlighting the potential for quantization and compression to enhance efficiency and performance.

\textbf{(2) Hashing and quantization models have benefits but limitations.} 
Rows 9 to 16 in Table \ref{tab:MainTable} indicate that methods like DHE \cite{DHE} and xLightFM \cite{xLightFM} reduce parameters significantly but degrade performance. This is due to their focus on compression without addressing data distribution imbalances caused by traditional quantization. These methods also require end-to-end training and lack portability. Our approach overcomes these issues by rebalancing quantized data distribution, achieving better performance.

\textbf{(3) Embedding lookup time correlates with embedding table size.}
Fig. \ref{fig:time_eff} shows that embedding lookup time is reduced by about 50\% with quantization due to smaller vocabulary sizes, indicating a strong link between inference time and memory size. Training latency analysis reveals that PQ quantization modifications add negligible latency, confirming the method's practicality. Full results are in Section 5.4. MEC's parameter compression and high-quality quantization enhance both inference efficiency and model performance.

\textbf{(4) MEC excels across all datasets.} 
Table 2 shows that $\text{MEC}_{GDCN}$ outperforms GDCN by 0.4\% on Avazu and 0.07\% on Criteo, with fewer parameters. As a pluggable model, MEC improves performance by up to 0.33\% and 0.41\% over PNN and GDCN, especially on Avazu due to its long-tailed distribution. This distribution often limits conventional models, but our method's popularity-based regularization enhances feature embedding quality.

    % \subsection{Industrial Dataset Performance}
    %     In Table \ref{tab:industry}, we present a comprehensive analysis of the performance and memory usage of MEC applied to an industrial dataset. 
    %     The comparison demonstrates that MEC, by utilizing Product Quantization's robust quantization capabilities reduces more than 99.7\% memory consumption of the embedding layer. 
    %     Additionally, our implementation of popularity-weighted regularization and contrastive learning effectively addresses common challenges encountered in quantized models, such as code allocation imbalance and uneven distribution of quantization centers. 
    %     Consequently, even with substantial parameter quantization, our method achieves performance that is comparable to, or even superior to, that of the baseline model.

% \subsection{Industrial Dataset Performance}

% Table \ref{tab:industry} shows that MEC reduces embedding layer memory usage by over 99.7\% using Product Quantization. The use of popularity-weighted regularization and contrastive learning addresses issues like code imbalance and uneven quantization distribution. Despite heavy quantization, our method matches or exceeds baseline model performance.

\subsection{Industrial Dataset Performance}

Table \ref{tab:industry} clearly demonstrates that MEC achieves over 99.7\% reduction in embedding layer memory usage through efficient Product Quantization. The integration of popularity-weighted regularization and contrastive learning effectively addresses typical challenges like code imbalance and uneven quantization distribution. Despite significant quantization, our method consistently maintains or even surpasses the performance of the baseline model.

    \begin{figure}[t]
        \centering
        \includegraphics[width=\textwidth]{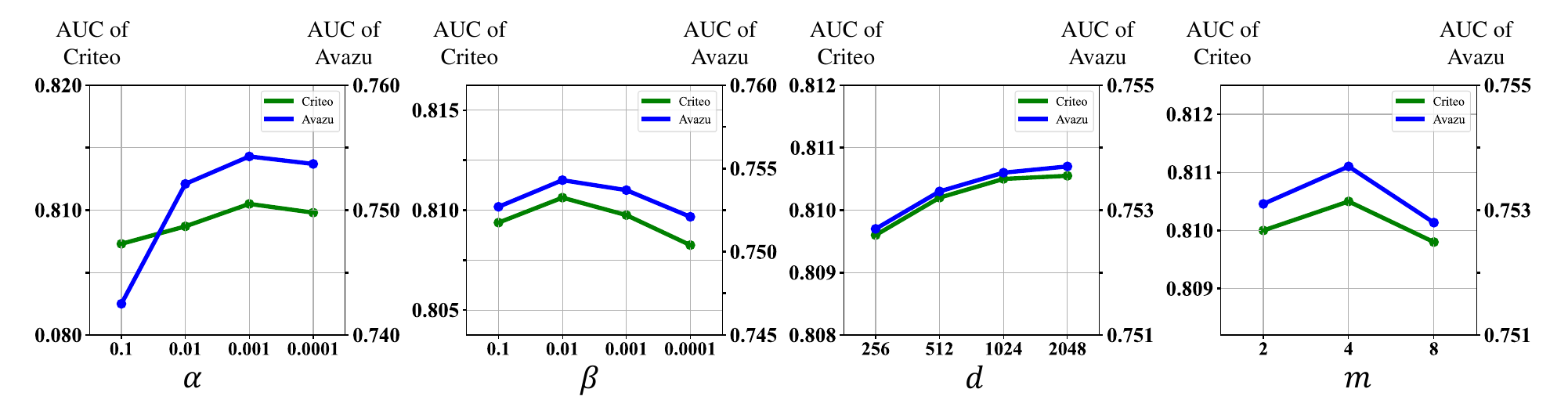}
        \vspace{-2em}
        \caption{Hyper-Parameter Performance}
    \label{fig:line_chart}
    \end{figure}
% \vspace{-3em}
    \subsection{In-depth Analysis}
        \input{Sections/Tables/industry}

\subsubsection{Analysis of Hyper-Parameter.}
As shown in Fig. \ref{fig:line_chart}, we analyzed key hyper-parameters in our MEC framework: the regularization loss coefficient ($\alpha$), contrastive loss coefficient ($\beta$), embedding dimension ($d$), and number of embedding layers ($m$). 
Optimal values were $\alpha = 0.001$, $\beta = 0.01$, $d = 2048$, and $m = 4$. Deviations from these values caused issues like underfitting, overfitting, information loss, or suboptimal relationship modeling.

\subsubsection{\textbf{Ablation Study of MEC}}
\input{Sections/Tables/ablation}

To thoroughly assess the contributions of different components in MEC, we evaluated three distinct variants on PNN \cite{PNN}: (1) without contrastive learning (w/o cons.); (2) without popularity-weighted regularization (w/o reg.); (3) using basic Product Quantization (PQ); and (4) frequency-based PQ (freq. PQ). The results, as shown in Table \ref{tab:Ablation}, were obtained using the Criteo and Avazu datasets.

% Applying PQ directly causes a significant performance drop. Although PQ reduces memory usage by quantizing the embedding table, it leads to suboptimal feature representation, weakening model expressiveness and CTR prediction accuracy.
% Including frequency information further decreases performance due to data imbalance. High-frequency features dominate the quantization centers, poorly representing low-frequency features and reducing accuracy.
% Regularization improves outcomes by preventing high-frequency features from dominating quantization, thus better representing low-frequency features. This counters the negative effects of frequency imbalance.
% Contrastive learning enhances separation between representations, promoting balanced code distribution. The combination of regularization and contrastive learning provides the best results by effectively balancing feature frequencies and improving quantized representation quality.

Directly applying PQ reduces performance by weakening feature representation and CTR prediction accuracy. Simply including frequency information exacerbates this by allowing high-frequency features to dominate, poorly representing low-frequency features. Regularization mitigates this imbalance by preventing domination and better representing low-frequency features. Contrastive learning enhances representation separation and code distribution. Together, regularization and contrastive learning balance feature frequencies and improve quantized representation quality.

\input{Sections/Tables/training_latency}

\subsubsection{Analysis of Training Latency}
\label{subsub:latency}
In this section, we analyze the training latency, given the computational constraints in real-world scenarios where training time needs careful consideration. 
Table \ref{tab:Latency} presents the results based on the Criteo and Avazu datasets, averaged over 10 runs. The PQ component adds negligible training time compared to the main CTR model. Popularity-weighted regularization does not increase time complexity, and while contrastive learning slightly increases training time, it remains manageable. Overall, our approach improves quantized embedding quality without significantly impacting training latency.

        % \subsubsection{\textbf{Analysis of Quantization Methods.}}
        %     \input{Sections/Tables/rq_vae}
        %     In this study, we evaluated various quantization methods for our MEC framework to balance memory efficiency and model performance. We compared Product Quantization (PQ) \cite{PQ} with Additive Quantization (AQ) \cite{AQ} and Residual Quantization (RQ) \cite{RQVAE}. Our experiments showed that PQ was the most effective among the three quantization methods.
            
        %     As shown in Table \ref{tab:rq_result}, RQ's performance lagged behind PQ due to high correlation among generated codes, causing instability in contrastive learning and unreliable negative samples. PQ, however, manages high-dimensional embedding tables by generating multiple sub-codebooks, preserving expressive power while reducing memory usage. The independence of these sub-codebooks ensures stability in contrastive learning, making PQ a robust choice for MEC.
        %     Similarly, AQ performed less effectively than PQ. AQ's additive process introduces redundancy and noise, reducing memory efficiency and stability. PQ avoids these issues with independent sub-codebooks, minimizing redundancy and optimizing memory use, thus enhancing contrastive learning stability.

\subsubsection{\textbf{Analysis of Quantization Methods}}

\input{Sections/Tables/rq_vae}

We evaluated quantization methods for MEC, comparing Product Quantization (PQ) with Additive Quantization (AQ) and Residual Quantization (RQ). Table \ref{tab:rq_result} shows that PQ outperforms the others. RQ suffers from code correlation, destabilizing contrastive learning, while AQ introduces redundancy and noise. PQ's independent sub-codebooks maintain stability and memory efficiency, making it the best choice for MEC.

\subsubsection{\textbf{Analysis of Pre-train Models}}
\input{Sections/Tables/pretrain_model_analysis}

We evaluated the impact of different pre-trained models (FM \cite{FM}, DeepFM \cite{DeepFM}, DCNv2 \cite{DCNv2}) on our framework for embedding generation and quantization, using PNN for downstream CTR prediction. Table \ref{tab:pretrain_model_analysis} shows minor performance variations across models, highlighting our framework's robust generalizability. This enables the use of simpler models in industrial applications without significant performance loss, allowing efficient deployment in various scenarios.

        % \subsubsection{\textbf{Additional Analyses}}
        %     We also conducted further analysis, including hyperparameter analysis in Appendix \ref{app:hyper} and analysis of pre-train models in Appendix \ref{app:pretrain_models}.To provide readers with an intuitive understanding of the uniformity in quantized embedding distributions, we present a visualization of code distribution in Appendix ~\ref{app:visual}. Additionally, the analysis of training latency is in Appendix \ref{app:latency}, providing deeper insights into the framework's time efficiency.

%% file: Sections/Tables/dataset.tex
\begin{table}[t]
    \caption{Dataset statistics}
    \centering
    \vspace{-0.2cm}
    \setlength{\tabcolsep}{1mm}
    \begin{tabular}{c|c|c|c}
        \toprule
        \textbf{Dataset} & \textbf{\#Interactions} & \textbf{\#Feature Fields} & \textbf{\#Features} \\
        \midrule
        Criteo & 45,840,617 & 39 & 1,086,810 \\
        Avazu & 40,428,968 & 22 & 2,018,012 \\
        \bottomrule
    \end{tabular}
    \label{tab:dataset}
    % \vspace{-10pt}
\end{table}

%% file: Sections/Tables/main_result.tex
\begin{table}[t]
    \centering
\setlength{\abovecaptionskip}{0.05cm}
\setlength{\belowcaptionskip}{0.2cm}
\caption{Overall performance comparison between the baselines and MEC instantiated on two competitive CTR prediction models across two datasets. 
% \gw{check the description} \kf{Sorry for this.}
The results table includes AUC, LogLoss, and memory usage for each model on the respective datasets. 
Bold values indicate a statistically significant level $p$-value<0.05 comparing MEC with the base model's performance in terms of AUC, LogLoss, and memory usage.}
\setlength{\tabcolsep}{2mm}{
\resizebox{0.9\textwidth}{!}{
\begin{tabular}{c|c|c|c|c|c|c}
\toprule
 &
    \multicolumn{3}{c|}{\multirow{-1}{*}{\textbf{Criteo}}} &
    \multicolumn{3}{c}{\multirow{-1}{*}{\textbf{Avazu}}} \\ \midrule
    \multicolumn{1}{c|}{\textbf{Model}} &
    \multicolumn{1}{c|}{\textbf{AUC}} &
    \multicolumn{1}{c|}{\textbf{Logloss}} &
    \multicolumn{1}{c|}{\textbf{Params}} &
    \multicolumn{1}{c|}{\textbf{AUC}} &
    \multicolumn{1}{c|}{\textbf{Logloss}} &
    \multicolumn{1}{c}{\textbf{Params}} \\   \midrule\midrule

    \textbf{LR} & 0.7879 & 0.4615 & - & 0.7514 & 0.3737 & - \\
    \textbf{FM} & 0.7945 & 0.4564 & 166.90MB & 0.7508 & 0.3755 & 103.91MB \\
    \textbf{AFM} & 0.8045 & 0.4475 & 166.92MB & 0.7541 & 0.3760 & 103.91MB \\
    \textbf{AFN} & 0.8055 & 0.4471 & 204.94MB & 0.7502 & 0.3764 & 141.86MB \\
    \textbf{SAM} & 0.8060 & 0.4490 & 167.13MB & 0.7528 & 0.3750 & 103.97MB \\
    \textbf{DeepFM} & 0.8069 & 0.4447 & 169.38MB & 0.7533 & 0.3744 & 105.60MB \\
    \textbf{AutoInt} & 0.8081 & 0.4433 & 169.96MB & 0.7515 & 0.3753 & 105.91MB \\
    \textbf{DCNv2} & 0.8099 & 0.4415 & 199.67MB & 0.7520 & 0.3746 & 116.37MB \\ \midrule
    \textbf{GDCN} & 0.8098 & 0.4417 & 172.86MB & 0.7507 & 0.3761 & 107.81MB \\
    \textbf{$\text{DHE}_{GDCN}$} & 0.8087 & 0.4427 & 8.089MB & 0.7505 & 0.3758 & 5.707MB \\
    \textbf{$\text{xLightFM}_{GDCN}$} & 0.8089 & 0.4422 & 7.659MB & 0.7504 & 0.3760 & 5.097MB \\
    \textbf{$\text{MEC}_{GDCN}$} & \textbf{0.8102} & \textbf{0.4415} & \textbf{7.649MB} & \textbf{0.7538} & \textbf{0.3728} & \textbf{4.567MB} \\ \midrule
    \textbf{PNN} & 0.8099 & 0.4418 & 168.80MB & 0.7531 & 0.3750 & 105.06MB \\
    \textbf{$\text{DHE}_{PNN}$} & 0.8088 & 0.4429 & 4.028MB & 0.7517 & 0.3763 & 2.957MB \\
    \textbf{$\text{xLightFM}_{PNN}$} & 0.8092 & 0.4423 & 3.598MB & 0.7524 & 0.3758 & 2.347MB \\
    \textbf{$\text{MEC}_{PNN}$} & \textbf{0.8105} & \textbf{0.4412} & \textbf{3.588MB} & \textbf{0.7556} & \textbf{0.3736} & \textbf{1.817MB} \\ \bottomrule

\end{tabular}
}}
\label{tab:MainTable}
% \vspace{-15pt}
\end{table}

%% file: Sections/Tables/industry.tex
\begin{table}[t]
\centering
\setlength{\abovecaptionskip}{0.05cm}
\setlength{\belowcaptionskip}{0.2cm}
\caption{Performance on industrial dataset}
\setlength{\tabcolsep}{2mm}{
\resizebox{0.75\textwidth}{!}{
\begin{tabular}{c|c|c|c|c|c}
\toprule
    \textbf{Metric} & \textbf{PRAUC}$\uparrow$ & \textbf{PCOC}$\uparrow$ & \textbf{AUC}$\uparrow$ & \textbf{LogLoss}$\downarrow$ & \textbf{Params}$\downarrow$ \\ \midrule
    \textbf{baseline} & 0.91913 & 0.96561 & 0.84141 & 0.40721 & 3142.3MB \\
    \textbf{MEC} & \textbf{0.91918} & \textbf{0.96673} & \textbf{0.84143} & \textbf{0.40683} & \textbf{7.014MB} \\
\bottomrule
\end{tabular}
}}
\label{tab:industry}
% \vspace{-10pt}
\end{table}

%% file: Sections/Tables/ablation.tex
\begin{table}[t]
\centering
\setlength{\abovecaptionskip}{0.05cm}
\setlength{\belowcaptionskip}{0.2cm}
\caption{Ablation study of MEC}
\setlength{\tabcolsep}{3mm}{
\resizebox{0.7\textwidth}{!}{
\begin{tabular}{c|c|c|c|c}
\toprule
 &
    \multicolumn{2}{c|}{\multirow{-1}{*}{\textbf{Criteo}}} &
    \multicolumn{2}{c}{\multirow{-1}{*}{\textbf{Avazu}}} \\ \midrule
    \multicolumn{1}{c|}{\textbf{Variants}} &
    \multicolumn{1}{c|}{\textbf{AUC}} &
    \multicolumn{1}{c|}{\textbf{Logloss}} &
    \multicolumn{1}{c|}{\textbf{AUC}} &
    \multicolumn{1}{c}{\textbf{Logloss}} \\   \midrule\midrule
    \textbf{$\text{MEC}_{PNN}$} & \textbf{0.8105} & \textbf{0.4412} & \textbf{0.7556} & \textbf{0.3736} \\
    \textbf{w/o cons.} & 0.8096 & 0.4418 & 0.7542 & 0.3739 \\
    \textbf{w/o reg.} & 0.8092 & 0.4421 & 0.7540 & 0.3741 \\
    \textbf{freq. PQ} & 0.8077 & 0.4451 & 0.7511 & 0.3755 \\
    \textbf{basic PQ} & 0.8084 & 0.4437 & 0.7528 & 0.3746 \\
    \textbf{PNN} & 0.8099 & 0.4418 & 0.7531 & 0.3750 \\
\bottomrule
\end{tabular}
}}
\label{tab:Ablation}
% \vspace{-10pt}
\end{table}

%% file: Sections/Tables/training_latency.tex
\begin{table}[t]
\centering
\setlength{\abovecaptionskip}{0.05cm}
\setlength{\belowcaptionskip}{0.2cm}
\caption{Latency of training stage}
\setlength{\tabcolsep}{3mm}{
\resizebox{0.45\textwidth}{!}{
\begin{tabular}{c|c|c}
\toprule
    \textbf{Metric} & \textbf{Criteo} & \textbf{Avazu} \\   \midrule\midrule
    \textbf{w/o PQ} & \textbf{24.07s} & \textbf{13.96s} \\
    \textbf{basic PQ} & 24.92s & 14.67s \\
    \textbf{w/o cons.} & 24.95s & 14.68s \\
    \textbf{w/o reg.} & 25.50s & 15.29s \\
    \textbf{$\text{MEC}_{PNN}$} & 25.56s & 15.31s \\
\bottomrule
\end{tabular}
}}
\label{tab:Latency}
% \vspace{-10pt}
\end{table}

%% file: Sections/Tables/rq_vae.tex
\begin{table}[t]
\centering
\setlength{\abovecaptionskip}{0.05cm}
\setlength{\belowcaptionskip}{0.2cm}
\caption{Comparison of quantization methods}
\setlength{\tabcolsep}{3mm}{
\resizebox{0.7\textwidth}{!}{
\begin{tabular}{c|c|c|c|c}
\toprule
 &
    \multicolumn{2}{c|}{\multirow{-1}{*}{\textbf{Criteo}}} &
    \multicolumn{2}{c}{\multirow{-1}{*}{\textbf{Avazu}}} \\ \midrule
    \multicolumn{1}{c|}{\textbf{Pretrain model}} &
    \multicolumn{1}{c|}{\textbf{AUC}} &
    \multicolumn{1}{c|}{\textbf{Logloss}} &
    \multicolumn{1}{c|}{\textbf{AUC}} &
    \multicolumn{1}{c}{\textbf{Logloss}} \\   \midrule\midrule
    \textbf{AQ+GDCN} & 0.8025 & 0.4507 & 0.7407 & 0.3871 \\
    \textbf{RQ+GDCN} & 0.8094 & 0.4421 & 0.7513 & 0.3752 \\
    \textbf{PQ+GDCN} & \textbf{0.8102} & \textbf{0.4415} & \textbf{0.7538} & \textbf{0.3728} \\  \midrule
    \textbf{AQ+PNN} & 0.8038 & 0.4480 & 0.7451 & 0.3813 \\
    \textbf{RQ+PNN} & 0.8099 & 0.4418 & 0.7528 & 0.3749 \\
    \textbf{PQ+PNN} & \textbf{0.8105} & \textbf{0.4412} & \textbf{0.7556} & \textbf{0.3737} \\
\bottomrule

\end{tabular}
}}
\label{tab:rq_result}
% \vspace{-14pt}
\end{table}

%% file: Sections/Tables/pretrain_model_analysis.tex
\begin{table}[t]
\centering
\setlength{\abovecaptionskip}{0.05cm}
\setlength{\belowcaptionskip}{0.2cm}
\caption{Performance on multiple pretrain models}
\setlength{\tabcolsep}{3mm}{
\resizebox{0.72\textwidth}{!}{
\begin{tabular}{c|c|c|c|c}
\toprule
 &
    \multicolumn{2}{c|}{\multirow{-1}{*}{\textbf{Criteo}}} &
    \multicolumn{2}{c}{\multirow{-1}{*}{\textbf{Avazu}}} \\ \midrule
    \multicolumn{1}{c|}{\textbf{Pretrain model}} &
    \multicolumn{1}{c|}{\textbf{AUC}} &
    \multicolumn{1}{c|}{\textbf{Logloss}} &
    \multicolumn{1}{c|}{\textbf{AUC}} &
    \multicolumn{1}{c}{\textbf{Logloss}} \\   \midrule\midrule
    \textbf{FM} & 0.8104 & 0.4413 & 0.7551 & 0.3741 \\
    \textbf{DeepFM} & \textbf{0.8105} & \textbf{0.4412} & \textbf{0.7556} & 0.3737 \\
    \textbf{DCNv2} & 0.8104 & 0.4412 & 0.7542 & \textbf{0.3736} \\
\bottomrule
\end{tabular}
}}
\label{tab:pretrain_model_analysis}
% \vspace{-13pt}
\end{table}

%% file: Sections/6_conclusion.tex
\section{Conclusion}
This paper tackles the memory consumption challenge in CTR prediction models caused by large embedding tables. We proposed a Model-agnostic Embedding Compression (MEC) framework, which combines popularity-weighted regularization (PWR) and contrastive learning to compress embeddings while maintaining high recommendation performance. Experiments on three real-world datasets demonstrate that MEC reduces memory usage by over 50x while achieving comparable or superior performance to state-of-the-art models. Our findings highlight MEC's potential for efficient and scalable CTR prediction.
\vspace{2pt}